\documentclass[conference]{IEEEtran}
\IEEEoverridecommandlockouts
\usepackage{cite}
\usepackage{amsmath,amssymb,amsfonts}
\usepackage{algorithmic}
\usepackage{graphicx}
\usepackage{textcomp}
\usepackage{xcolor}
\usepackage{multirow}
\usepackage{multicol}
\usepackage{subcaption}
\usepackage{lipsum}
\usepackage{stfloats}
\usepackage{booktabs}
\usepackage{tabularx}
\usepackage{ragged2e}
\usepackage[hidelinks]{hyperref}
\usepackage{array}
\newcolumntype{M}[1]{>{\centering\arraybackslash}m{#1}}
\usepackage{balance}
\usepackage{bm}
\usepackage{tikz}

\DeclareCaptionStyle{centeredlabel}{
  justification=centering,
  labelsep=newline,
  labelfont=normalfont,
  textfont=small
}

\def\BibTeX{{\rm B\kern-.05em{\sc i\kern-.025em b}\kern-.08em
    T\kern-.1667em\lower.7ex\hbox{E}\kern-.125emX}}
    
\begin{document}

\pagestyle{plain}

\title{ProbSplat: Efficient Probabilistic Hardware for Gaussian Splatting in 3D Scene Reconstruction\\
% {\footnotesize \textsuperscript{*}Note: Sub-titles are not captured in Xplore and
% should not be used}
% \thanks{Identify applicable funding agency here. If none, delete this.}
}

% \author{Anonymous authors}

\author{Siddarth Gottumukkula$^{*1}$, M P Samartha$^{*1}$, Vedant Pahariya$^1$, Priyanshi Jain$^1$, \\ Amit Ranjan Trivedi$^2$, \textit{Senior Member, IEEE}, and Priyesh Shukla$^1$, \textit{Member, IEEE}\\
$^1$International Institute of Information Technology, Hyderabad, India\\ 
$^2$University of Illinois, Chicago, USA,
email: priyesh.shukla@iiit.ac.in}

% \author{\IEEEauthorblockN{Siddarth Gottumukkula}
% \IEEEauthorblockA{\textit{dept. name of organization (of Aff.)} \\
% \textit{name of organization (of Aff.)}}
% \and
% \IEEEauthorblockN{M P Samartha}
% \IEEEauthorblockA{\textit{dept. name of organization (of Aff.)} \\
% \textit{name of organization (of Aff.)}}
% \and
% \IEEEauthorblockN{Priyesh Shukla}
% \IEEEauthorblockA{\textit{dept. name of organization (of Aff.)} \\
% \textit{name of organization (of Aff.)}}
% \and
% \IEEEauthorblockN{Vedant Pahariya}
% \IEEEauthorblockA{\textit{dept. name of organization (of Aff.)} \\
% \textit{name of organization (of Aff.)}}
% \and
% \IEEEauthorblockN{Priyanshi Jain}
% \IEEEauthorblockA{\textit{dept. name of organization (of Aff.)} \\
% \textit{name of organization (of Aff.)}}
% \and
% \IEEEauthorblockN{6\textsuperscript{th} Given Name Surname}
% \IEEEauthorblockA{\textit{dept. name of organization (of Aff.)} \\
% \textit{name of organization (of Aff.)}}
% }

\maketitle

\begin{tikzpicture}[remember picture,overlay]
\node[anchor=south, yshift=10pt] at (current page.south) {
    \parbox{\textwidth}{
        \centering \footnotesize 
        \textcopyright~2026 IEEE. Personal use of this material is permitted. Permission from IEEE must be obtained for all other uses, in any current or future media, including reprinting/republishing this material for advertising or promotional purposes, creating new collective works, for resale or redistribution to servers or lists, or reuse of any copyrighted component of this work in other works.\\ 
        Accepted for publication in the 2026 IEEE Computer Society Annual Symposium on VLSI (ISVLSI).
    }
};
\end{tikzpicture}
\begin{abstract}

This paper presents ProbSplat, a Compute-in-Memory (CIM)-inspired architecture based on programmable and energy efficient floating-gate inverter columns for probabilistic computing. Improving upon our prior work, ProbSplat programs and stores both means and variances of Gaussian mixture components, and evaluates log-likelihood for gaussian splatting during scene reconstruction with high energy efficiency, suitable for robotics and augmented/virtual reality (AR/VR) at the edge. Our proposed scheme enables independent control of both mean and variance via deterministic adjustment of floating-gate MOSFET threshold voltages, increasing the fidelity of hardware to program probability distributions. The design is simulated in 180nm CMOS on 1.8 V at 50 MHz and achieves mean–variance independence with $<$ 2.4\% deviation during 3-D Gaussian mixture modeling. Compared to conventional digital implementations, ProbSplat significantly reduces compute complexity, memory footprint, and power consumption. The scalable framework consumes 18pJ energy per log-likelihood inference with 4-bit precision while operating for 500 mixture functions in a 3-D GMM. Scene reconstruction with ProbSplat's characteristics gave satisfactory fidelity of 21.99 PSNR (dB) at 8-bit precision.
\end{abstract}

\begin{IEEEkeywords}
Gaussian splatting, In-memory Probabilistic Computing, Programmable floating-gate inverters
\end{IEEEkeywords}

\section{Introduction}
Probabilistic modeling and computation are at the core of scientific computing~\cite{probabilistic-approx},~\cite{probabilistic-photonics}, machine learning~\cite{MLprobabilistic}, robotics~\cite{robotics-localization}, and AR/VR~\cite{arvr_4dgs}. Probabilistic models have critical use cases for estimating uncertainty in machine learning predictions for real-time tasks such as self-driving cars~\cite{Autonomous_driving} and surgical robots~\cite{surgical_gmm} to make safe and informed decisions. In 3-D vision applications such as AR/VR and drone navigation, probabilistic models have vital and extensive use for scene reconstruction~\cite{Charatan_2024_CVPR}, localization~\cite{Matsuki_2024_CVPR}, and mapping~\cite{mapping_survey}.

Conventional point cloud~\cite{fu2024colmap} based 3-D vision processing, such as scene rendering and mapping, demands an enormous memory footprint to store millions of point cloud coordinates. Consequently, local (edge) real-time processing of point clouds is intractable and extremely power hungry. Alternatively, representing the scene with a probability distribution, such as the mixture of Gaussians (GMMs)~\cite{gmm}, requires storing relevant parameters, such as means and variances, drastically reducing the memory footprint and power demand.

\begin{figure}[t]
    \centering
    \begin{subfigure}{0.49\textwidth}
        \centering
        \includegraphics[width=0.975\linewidth]{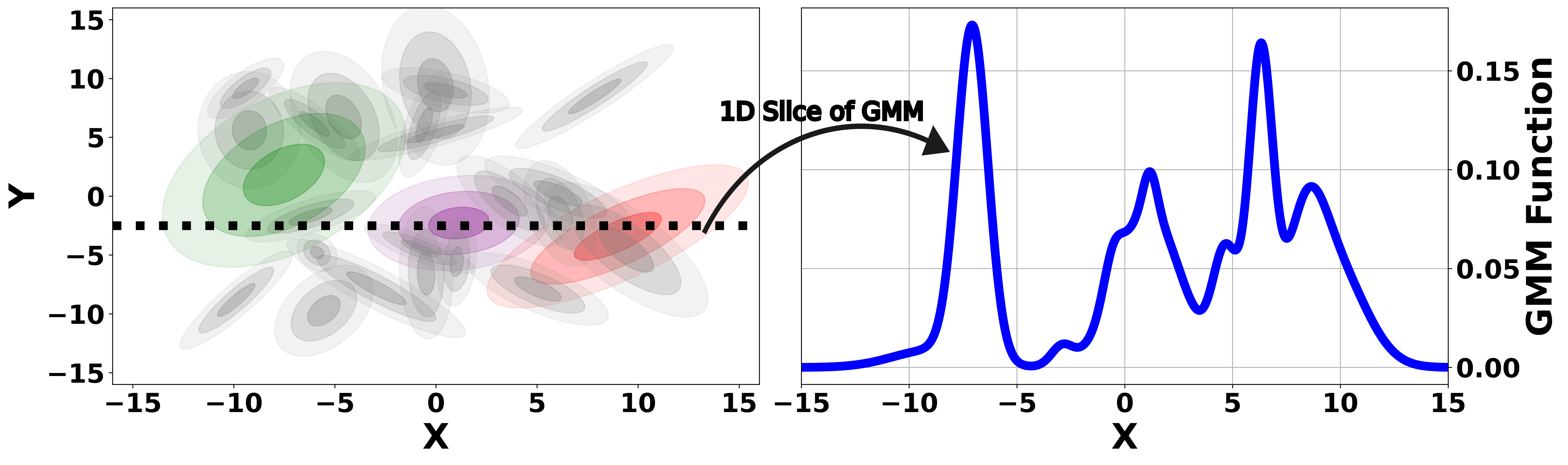}
        \caption*{(a)}
    \end{subfigure}
    \hfill
 \begin{subfigure}{0.49\textwidth}
        \centering
        \includegraphics[width=0.975\linewidth]{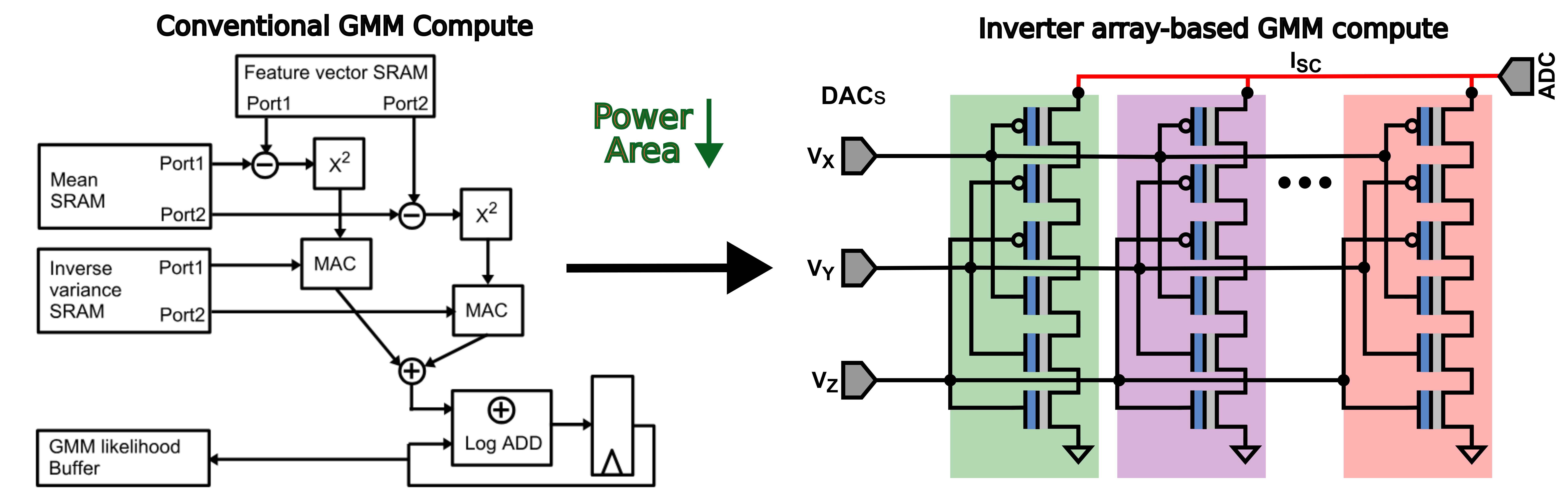}
        \caption*{(b)}
    \end{subfigure}
    \caption{\small{(a) Gaussian Mixture Model (GMM) of a scene (left) and 1-D representation (right). (b) Left: Conventional digital implementation of GMM using memory, MAC and other units. Right: Proposed architecture to compute GMM using 6T Floating-Gate Inverter Arrays.} 
    \vspace{-2em}}
    \label{Top Fig}
\end{figure}

Computing the probability distribution and associated metrics, such as log-likelihood estimates, is silicon area and power-intensive, requiring complex circuits involving adders, multipliers, and look-up tables for exponential/logarithmic functions, as shown in Fig. 1(b) (left). Compute-in-memory (CIM)-inspired circuits and architectures~\cite{cim},~\cite{cim-survey}, on the contrary, are promisingly area-power efficient since the storage and compute are co-located, alleviating energy-heavy dataflow between memory and compute. Prior work~\cite{6t-inv} encoded the mean of a multivariate Gaussian-like distribution into the threshold voltages of Floating Gate MOSFETs. The map was represented using the harmonic mean of a Gaussian-like mixture model (HMGM) emulated by the 6T inverter array, as shown in Fig. 1(b) (right). While the prior approach allows tuning the mean, the variance remains fixed, limiting the scope of scene representation using probability distributions. Complex scenes require multivariate distributions of varying variances. In this work, we further investigate and present a method to controllably tune the variance independent of the mean on a 6T-inverter array for applications such as gaussian splatting~\cite{3dgs} in 3D scene reconstruction.

The paper is organized as: Section II presents 6T floating-gate inverter array for probabilistic computing. Section III presents detailed analyses of GMM parameter programming on hardware. Section IV shows energy-performance characterization of ProbSplat, and Section V concludes the paper. 

\begin{figure*}[t]
  \centering
  \setlength{\abovecaptionskip}{-1pt}
  % First row
  \begin{subfigure}{\textwidth}
    \centering
    \includegraphics[width=0.8\linewidth]{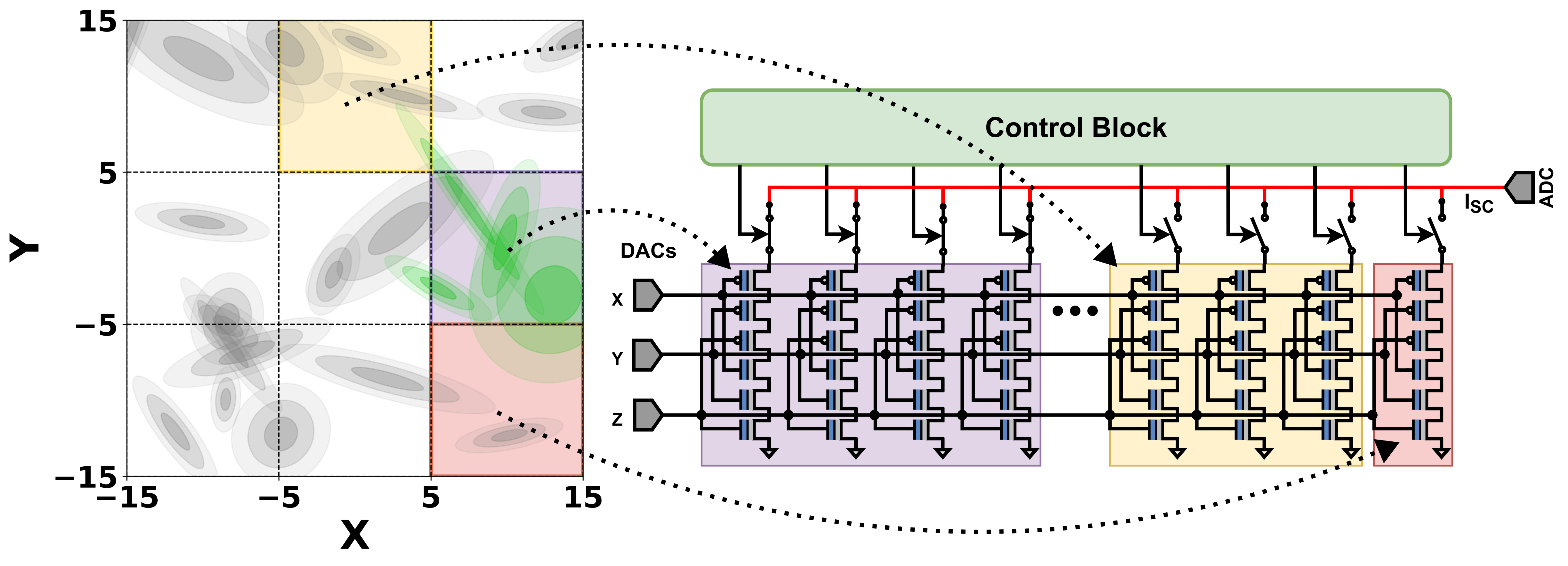}
    \caption*{(a)}
  \end{subfigure}
  \hfill
  \begin{subfigure}{0.3\textwidth}
    \centering
    \raisebox{1.5mm}{ % Adjust vertical alignment if needed
    \includegraphics[width=\linewidth]{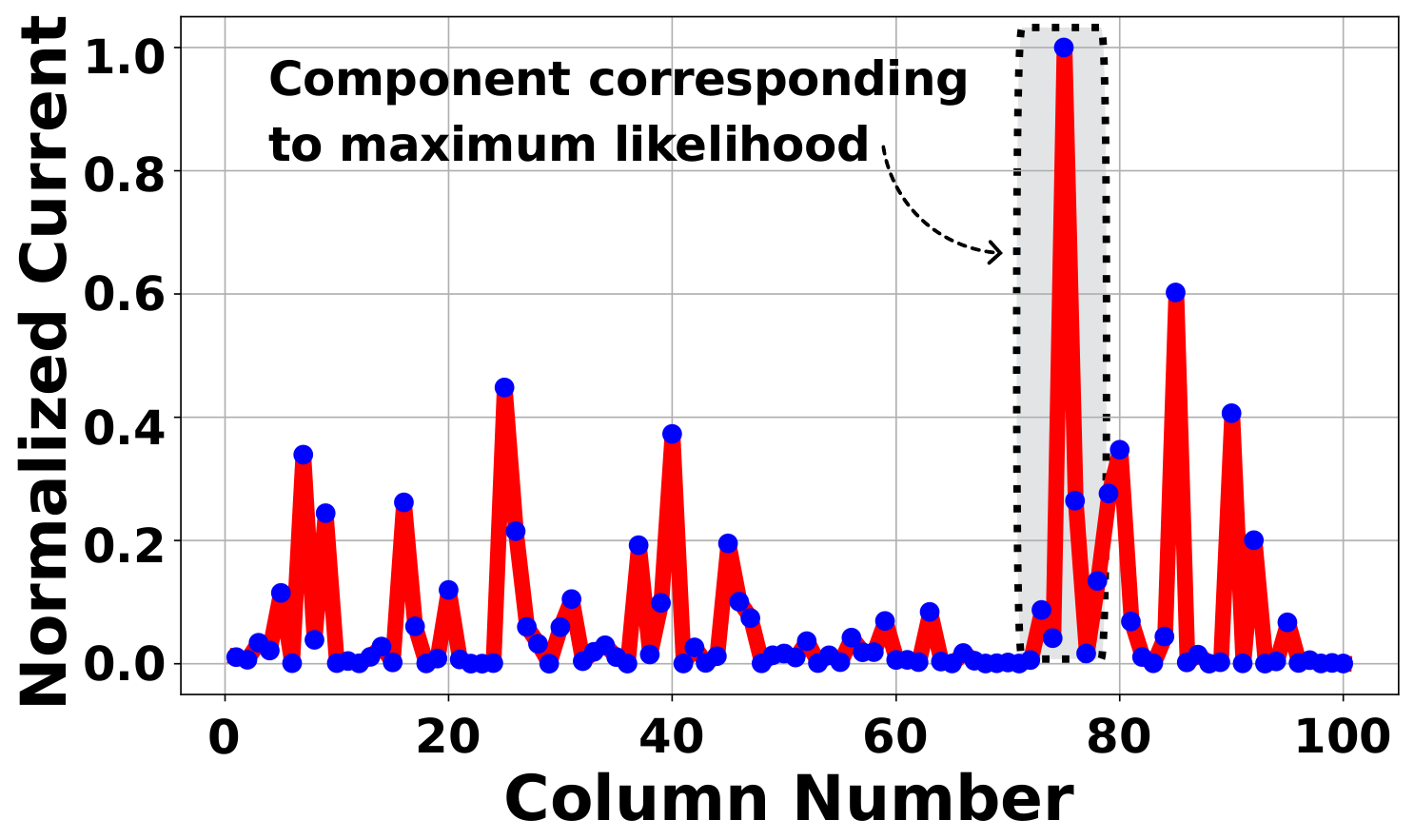}
    }
    \caption*{(b)}
  \end{subfigure}
  % \hfill
  \begin{subfigure}{0.3\textwidth}
    \centering
    \raisebox{1.5mm}{ % Adjust vertical alignment if needed
      \includegraphics[width=\linewidth]{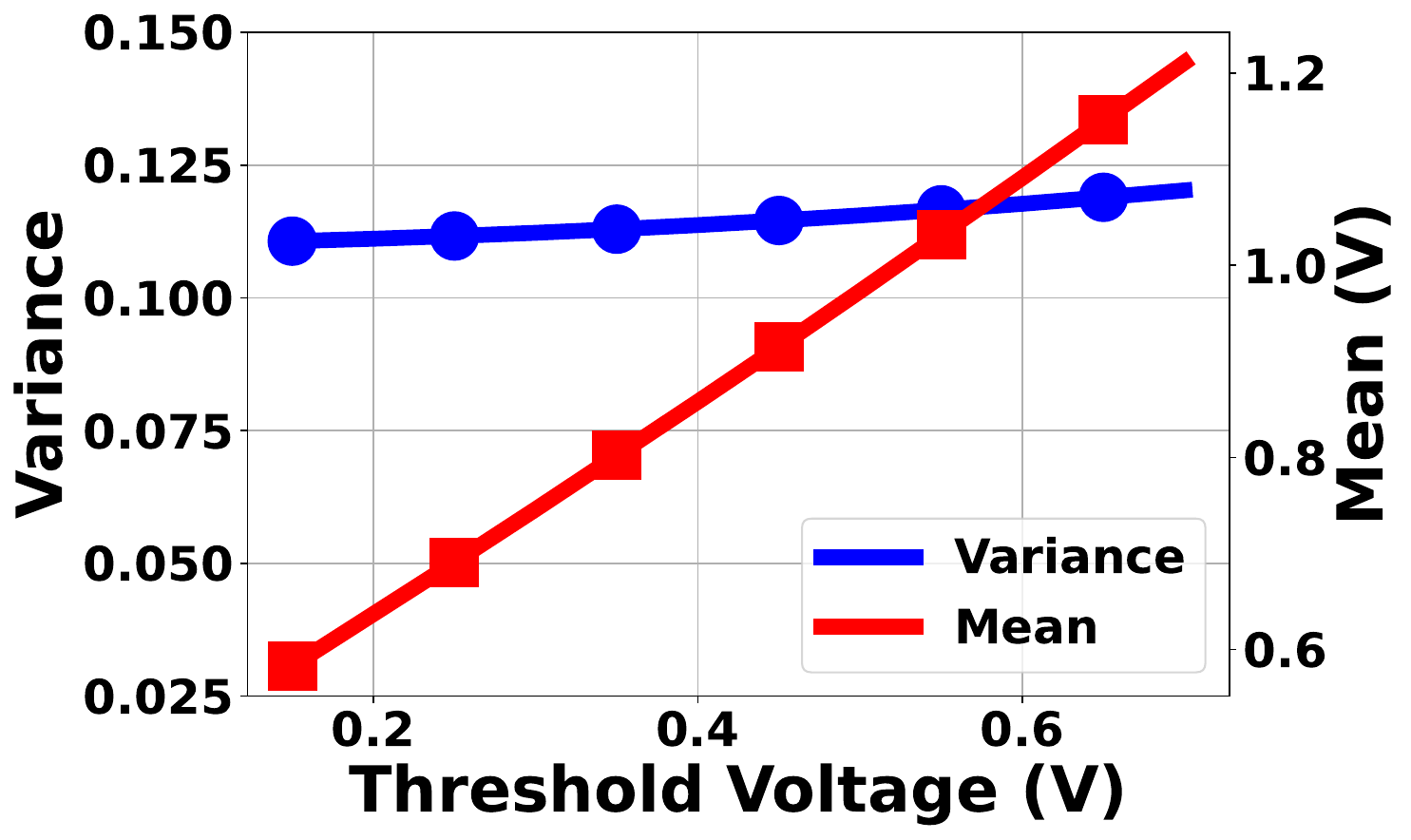}
    }
    \caption*{(c)}
  \end{subfigure}
  % \hfill
  \begin{subfigure}{0.3\textwidth}
    \centering
    \raisebox{1.5mm}{ % Adjust vertical alignment if needed
      \includegraphics[width=\linewidth]{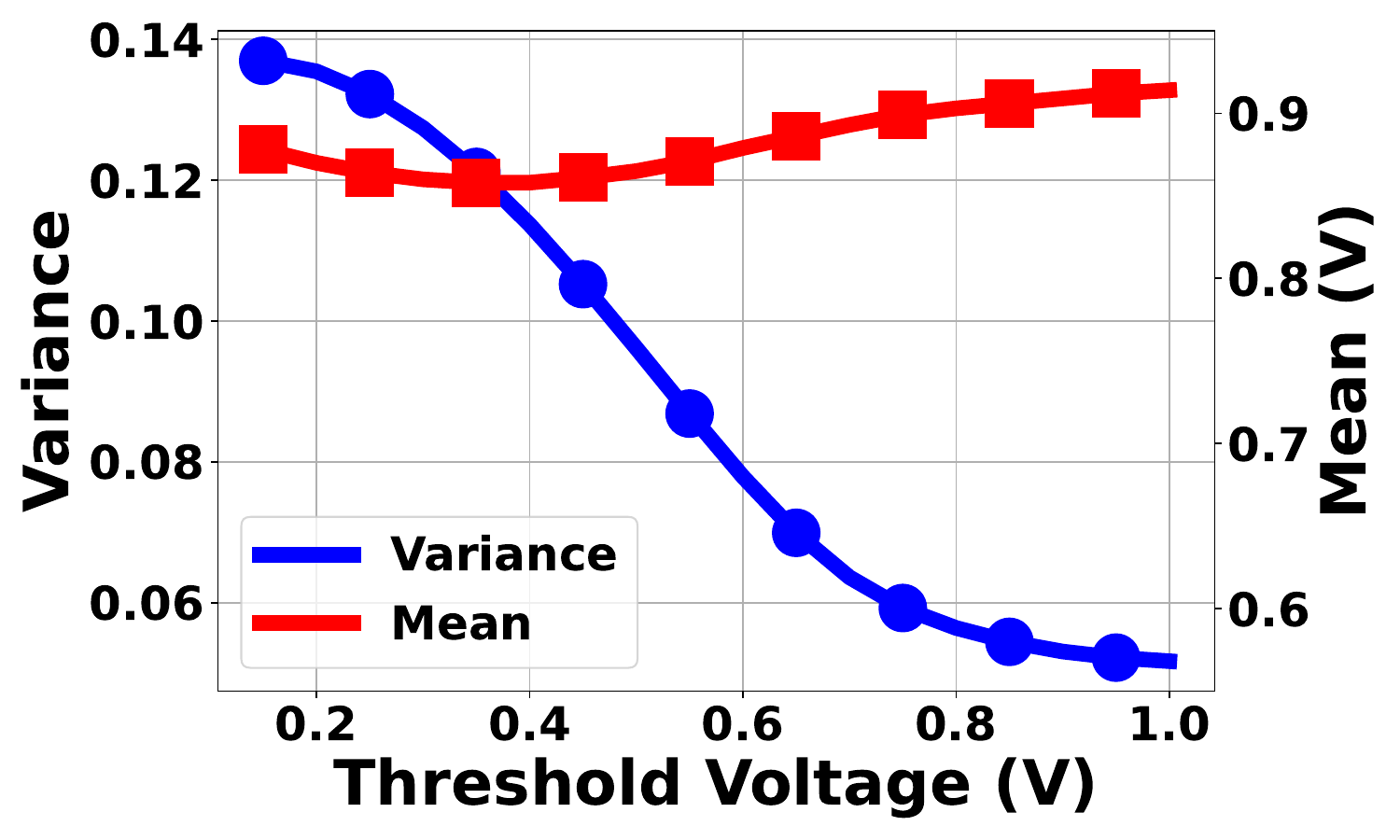}
    }
    \caption*{(d)}
  \end{subfigure}
    \caption{\small{(a) A scene represented by gaussian mixture components with varying means and covariances mapped to a 6T-inverter columns. The current through each column is the estimated log-likelihood over GMM. The control signals activate gaussian components clustered around the evaluated location. The transmission gate structure reduces power consumption. (b) Current distribution through each column for a particular input for a 100-column simulation. (c - d) show the change in variance when the mean is changed and vice versa. The average deviation in mean and variance is 2.37\% and 2.17\%.}}
    \label{application array}
    \vspace{-12pt}
\end{figure*}

\section{Probabilistic Inverter Array and Applications}

The traditional Von Neumann architecture poses a significant memory bandwidth bottleneck with growing data and parameter size. Compute-in-Memory techniques tackle this problem by exploiting the circuit characteristics to perform both storage and computation to get the desired results. The 6T-Inverter proposed in~\cite{6t-inv} stores a 3-D Gaussian's mean in Floating Gate (FG) MOSFETs by programming the threshold voltages. However, the variance was constant, under-utilizing the architecture's potential. We further enhanced the method to change the threshold voltages of the FG-MOSFETs in deterministic ways to tune the variance independent of the mean.

\captionsetup[table]{style=default}

The mean changes by $\Delta$ upon setting $V_{T_n}' = V_{T_no} + \Delta$ and $|V_{T_p}|' = |V_{T_p}| - \Delta$. To change the variance, both $V_{T_n}$ and $|V_{T_p}|$ must be changed by the same amount. This changes the spread of $I_{SC}$, the short-circuit current. 

The 6T inverter array as a CIM architecture as shown in Fig. \ref{application array}(a) has several use cases. For instance, the locality of a drone can be inferred by representing the scene as a mixture of Gaussians and evaluating the likelihood of the drone. The resulting likelihood current of inverter columns would be high if the drone is inside a particular Gaussian. Using a CIM block where 6T-inverter columns are connected in parallel, we can model each Gaussian of the scene by an inverter column. This would help us determine whether the drone is in the area envelope or outside it, as seen in Fig. \ref{application array}(b). The scheme can be used to model obstacles in a scene using Gaussians, represented by columns of inverters. The column current rises as the drone approaches obstacles, and corrective action can be taken accordingly.

\begin{figure}[t]
    \centering
    \setlength{\belowcaptionskip}{-5pt}
    \begin{subfigure}[b]{0.23\textwidth}
        \centering
        \includegraphics[width=\textwidth]{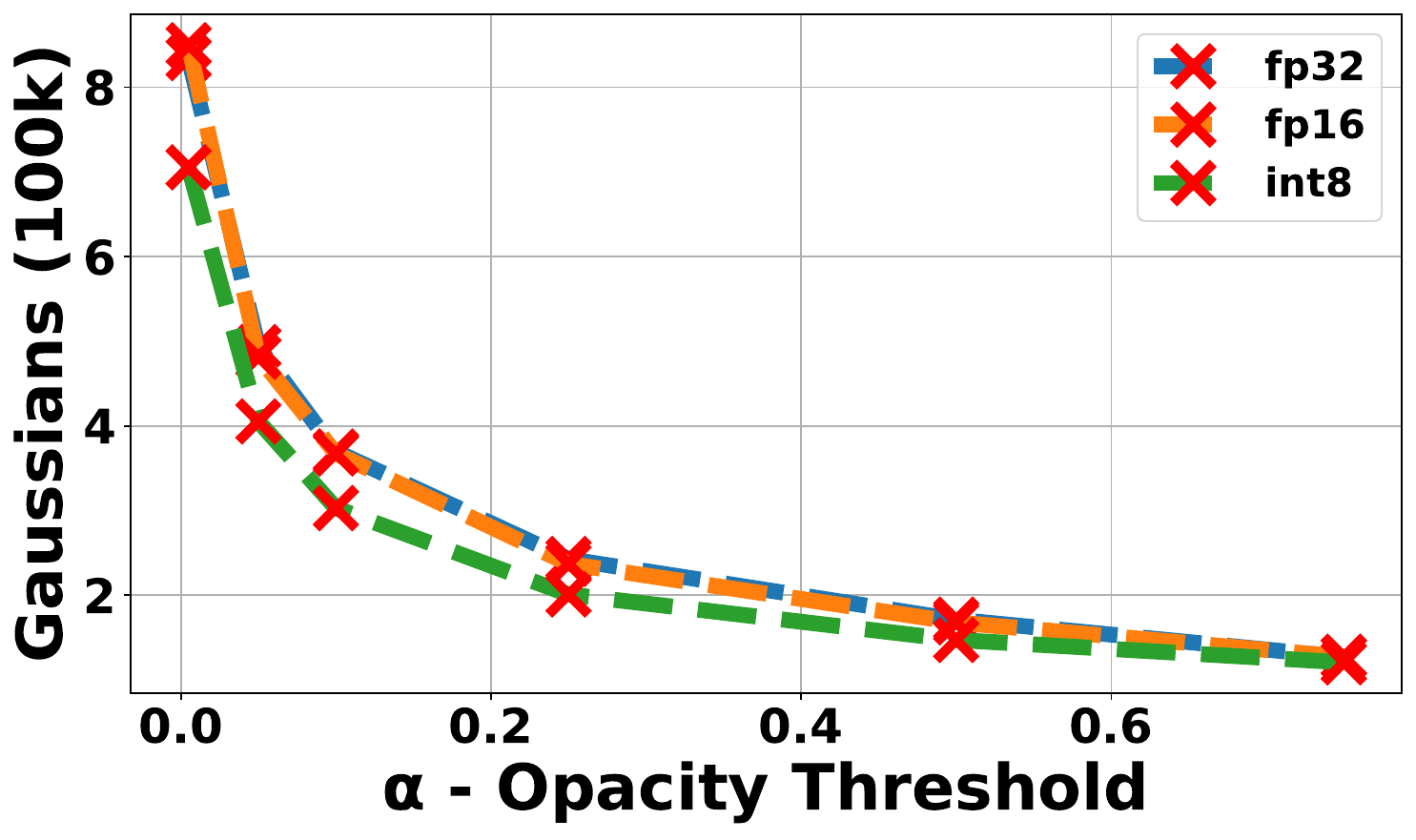} % Replace with your first plot
        \caption{}
        % \label{}
    \end{subfigure}
    \hfill % This command inserts a horizontal space between the two subfigures
    \begin{subfigure}[b]{0.23\textwidth}
        \centering
        \includegraphics[width=\textwidth]{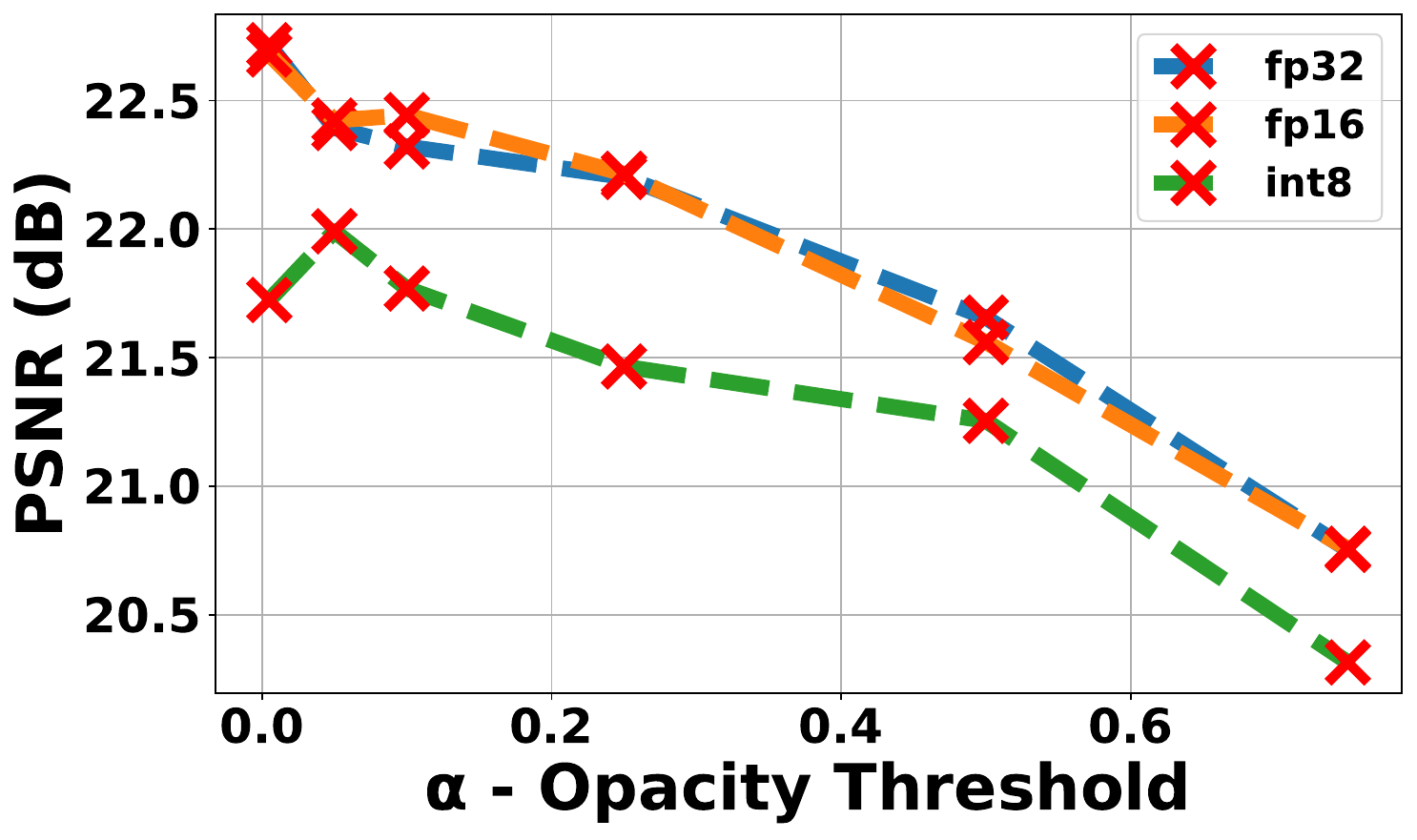} % Replace with your second plot
        \caption{}
        % \label{fig:plotB}
    \end{subfigure}
    \caption{\small{(a) Gaussian components below the opacity threshold are filtered out, reducing the number of required inverter columns. (b) Effect of changing the opacity threshold on the PSNR of the rendered image over different quantization schemes.}}
    % \vspace{-1em}
    \label{Quantization Results}
\end{figure}

Another promising application benefitting from proposed architecture is Gaussian Splatting~\cite{3dgs}, widely used in 3D scene reconstruction. At 32-bit parameter precision, the threshold voltage range of an FG-MOSFET requires $2^{32}$ quantization bins. This requires a very high precision ADC to read the current changes. Hence, low bit quantization is essential for these architectures. From Fig. \ref{Quantization Results}(b), we see that the PSNR is not affected much by 16-bit quantization but drops significantly with 8-bit quantization. We can map the means and the variances of the gaussians representing a scene to the inverter columns to reduce the total memory footprint. Typical Gaussian Splatting scene representations consist of about 800,000 gaussians. These can be pruned further by setting an opacity threshold (Fig. \ref{Quantization Results}(a)), ultimately producing a sizable decrease in the occupied memory and computations. We subsequently discuss the impact of 6T inverter column-based gaussian splatting for scene reconstruction as shown in Fig. \ref{fig:scenes}.

% Another application of the proposed architecture in image processing is Gaussian blurring~\cite{gaussian_blurr}. Using a 4T-inverter, we can store the pixel coordinates in the thresholds of the transistors. For each color channel, we would have an array of inverter columns. The value of each pixel will be mapped to the number of columns assigned to the pixel. The variances of the inverter columns characterize the gaussian kernel. By applying the location of a pixel as the input voltage, the cumulative current is read, and we get a Gaussian-blurred new pixel value as the output. Ultimately, we can obtain a resultant Gaussian-blurred image output by sweeping the input voltages over all pixels. 

\section{Parameter Programming and Analyses}

\begin{figure}[h]
    \centering
    \setlength{\belowcaptionskip}{-5pt}
    \begin{subfigure}[b]{0.23\textwidth}
        \centering
        \includegraphics[width=\textwidth]{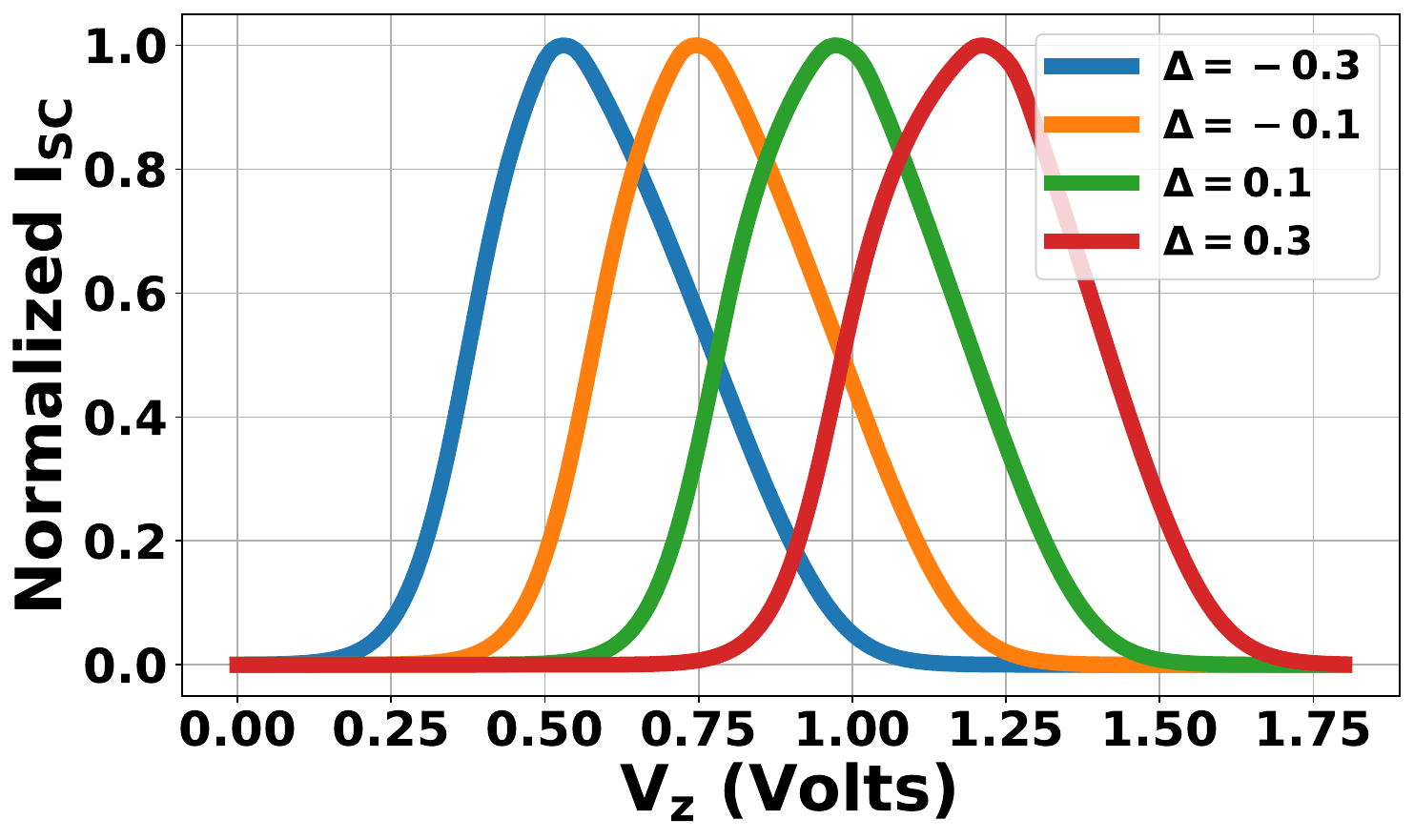} % Replace with your first plot
        \caption{}
        % \label{}
    \end{subfigure}
    \hfill % This command inserts a horizontal space between the two subfigures
    \begin{subfigure}[b]{0.23\textwidth}
        \centering
        \includegraphics[width=\textwidth]{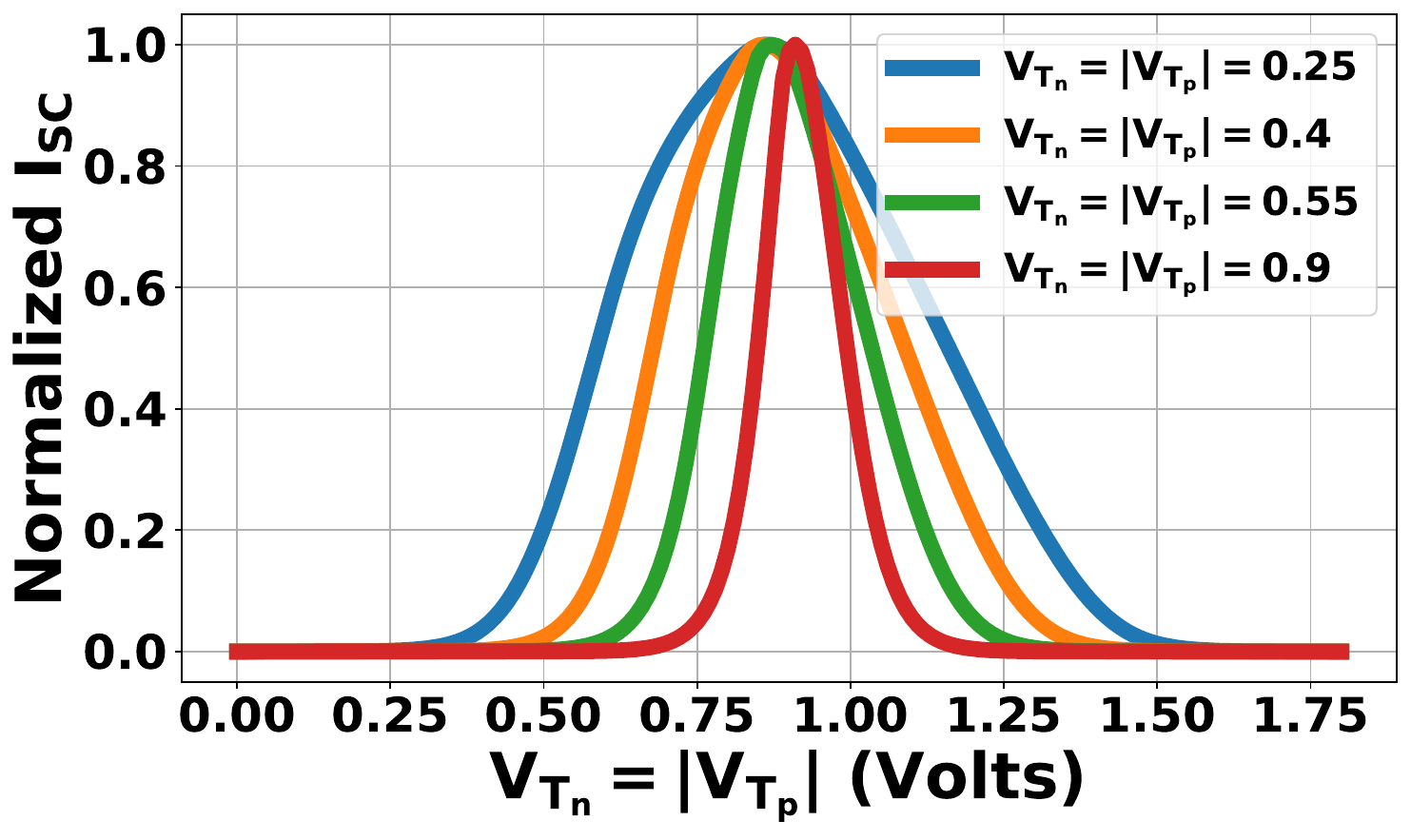} % Replace with your second plot
        \caption{}
        % \label{fig:plotB}
    \end{subfigure}
    \caption{\small{(a) Programming the means where $V_{T_n} = 0.4 +\Delta $, $|V_{T_p}| = 0.4 -\Delta$ and $V_x = V_y = 0.9 + \Delta$. (b) Programming variance with $V_x = V_y = 0.9V$. The results have been recorded with $W_p = 3*W_n$.}}
    \label{3 means and vars}
\end{figure}

All the simulations have been carried out using the TSMC 180nm technology node using $V_{DD} = 1.8V$. While gaussian transistors~\cite{gaussian-transistors} could yield superior results, we employ CMOS technology for its accessibility and ease of use to analyze the characteristics qualitatively. Fig. \ref{3 means and vars} captures different $I_{SC}$ curves while changing the means and variances.

\begin{figure}[t!]
    \centering
    \setlength{\belowcaptionskip}{-5pt}
    \begin{subfigure}[b]{0.23\textwidth}
        \centering
        \includegraphics[width=\textwidth]{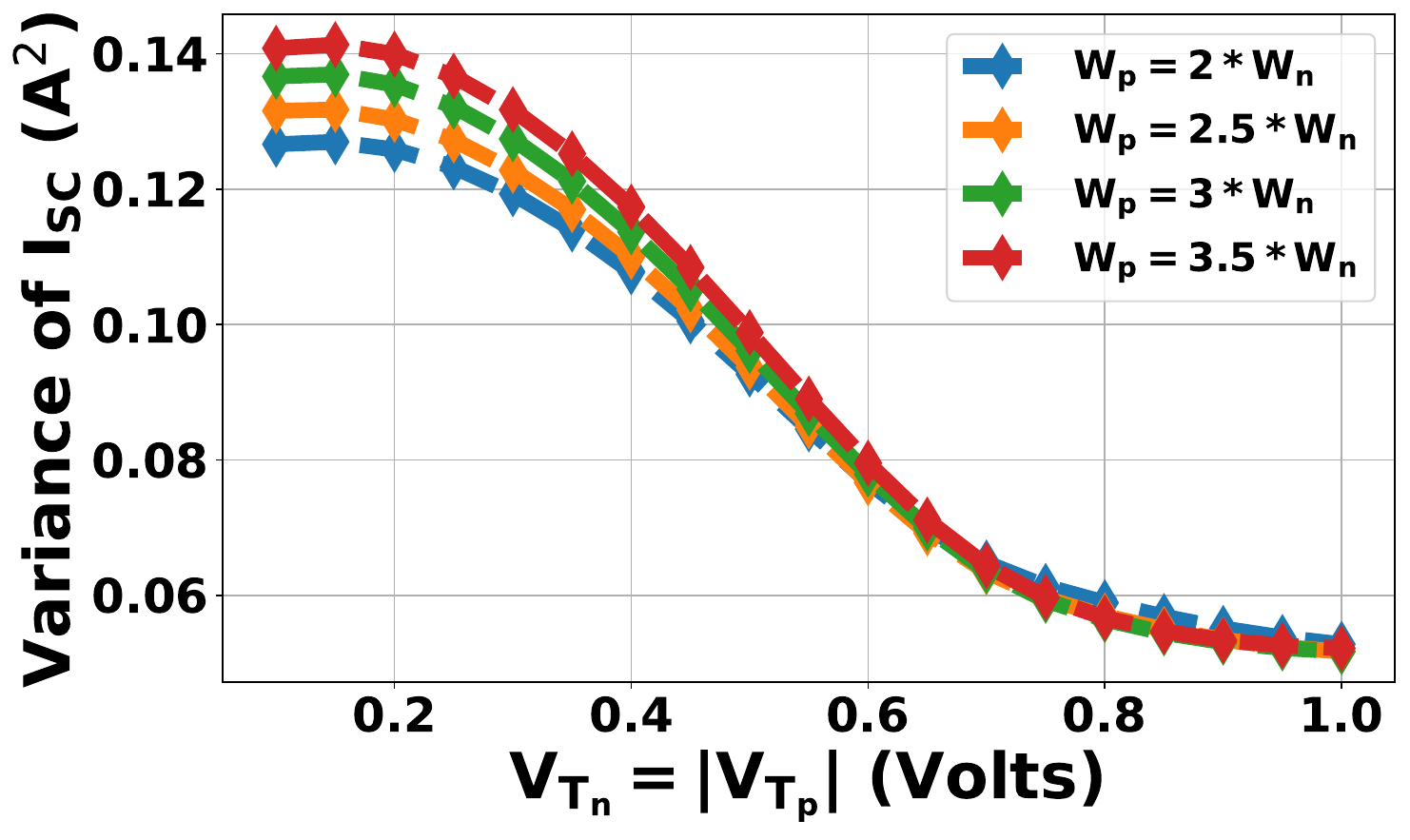} % Replace with your first plot
        \caption{}
        % \label{}
    \end{subfigure}
    \hfill % This command inserts a horizontal space between the two subfigures
    \begin{subfigure}[b]{0.23\textwidth}
        \centering
        \includegraphics[width=\textwidth]{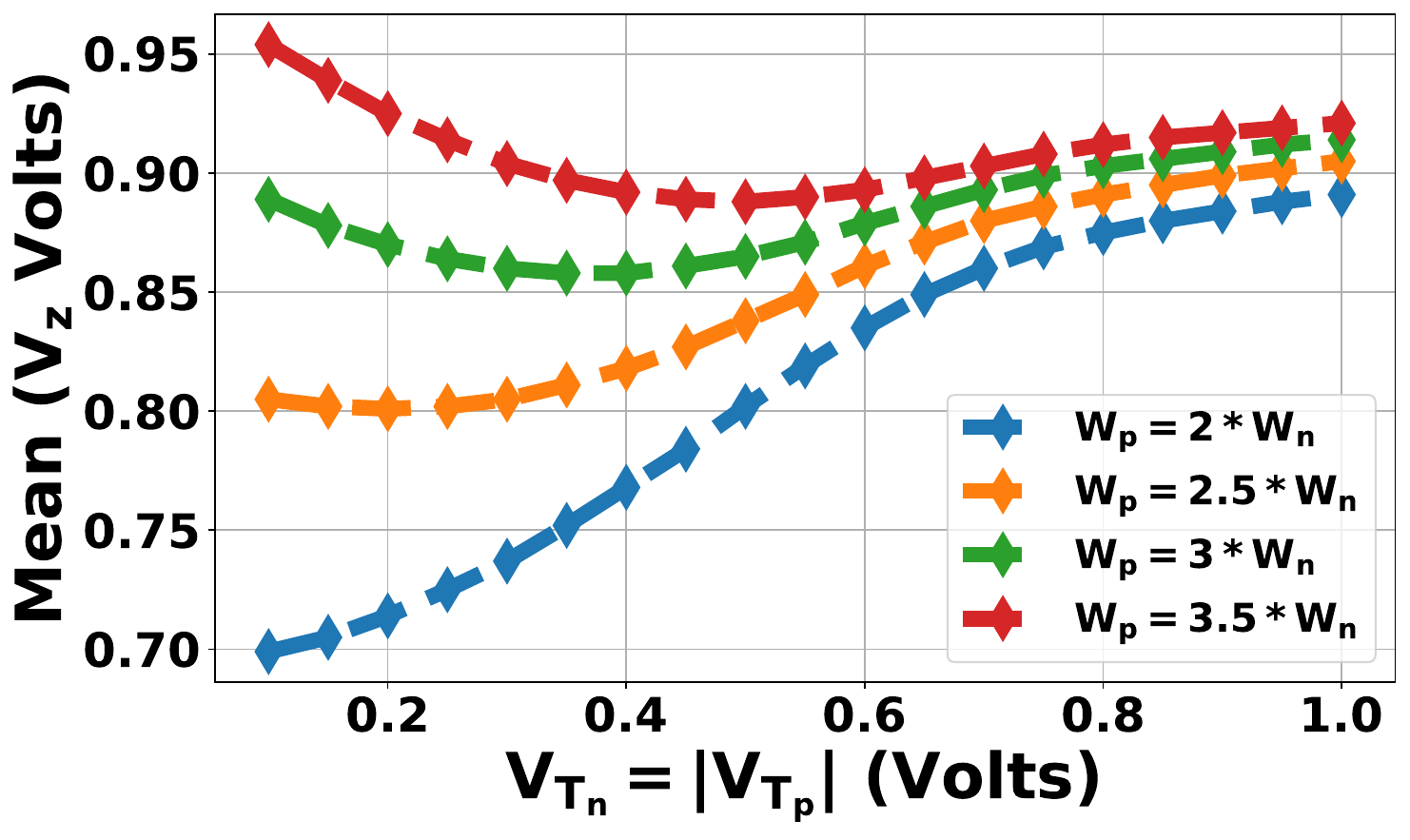} % Replace with your second plot
        \caption{}
        % \label{fig:plotB}
    \end{subfigure}

    % \begin{subfigure}[b]{0.24\textwidth}
    %     \centering
    %     \includegraphics[width=\textwidth]{Plots/means_and_vars/curr_new_main.pdf} % Replace with your second plot
    %     \caption{}
    %     % \label{fig:plotB}
    % \end{subfigure}
    
    \caption{\small{(a) Changes in variance with threshold voltage for different $W_p:W_n$ ratios. (b) The change in the mean upon change in the variance. $W_p = 3*W_n$ gives the least deviation in the mean.}}
    \label{main plots}
    \vspace{-8pt}
\end{figure}

\begin{figure*}[t!] 
    \centering
    \setlength{\abovecaptionskip}{-3pt}
    \setlength{\belowcaptionskip}{-5pt}
    \setlength{\tabcolsep}{0pt} % horizontal padding between columns
    \renewcommand{\arraystretch}{0.95} % vertical padding between rows
    \newcommand{\img}{\includegraphics[width=0.20\textwidth]{example-image-a}}
    \newlength{\imgheight}
    \settoheight{\imgheight}{\img}

    \begin{tabular}{m{0.05\textwidth} c @{\hspace{2pt}} c @{\hspace{10pt}} c @{\hspace{2pt}} c}

    \raisebox{1.2\height}{\rotatebox{90}{\parbox{\imgheight}{\centering\textbf{Ground Truth}}}} & 
    \multicolumn{2}{c @{\hspace{10pt}}}{\includegraphics[width=0.36\textwidth]{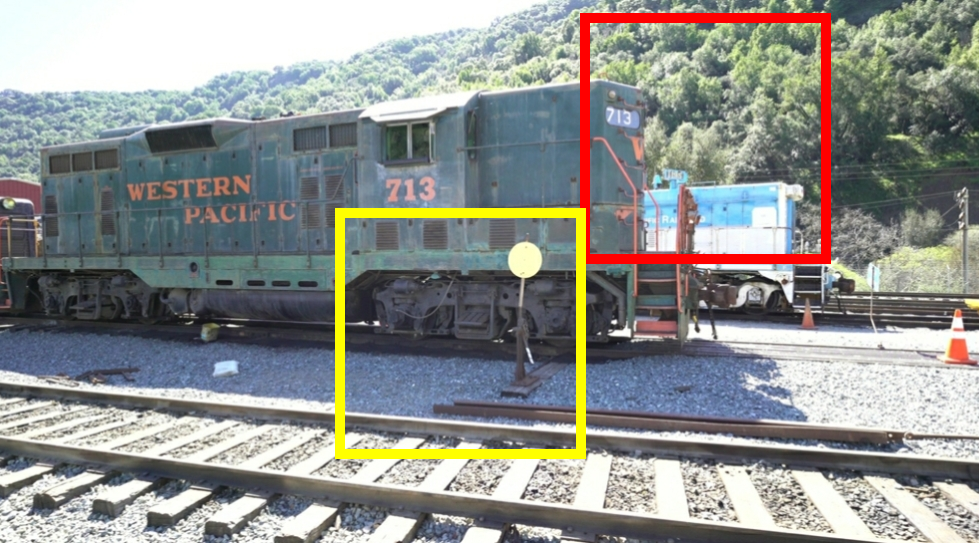}} & 
    \multicolumn{2}{c}{\includegraphics[width=0.36\textwidth]{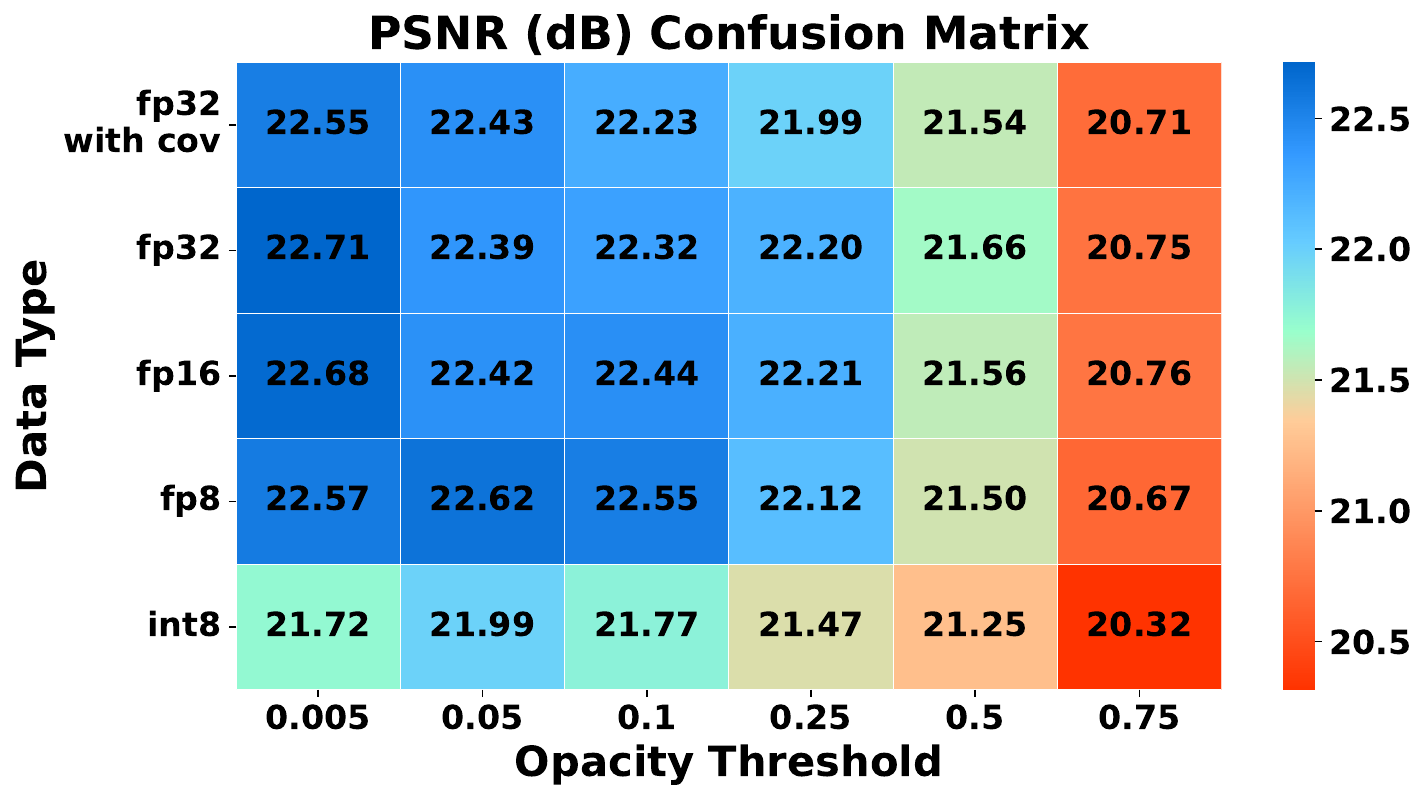}} \\[-80pt]

    \raisebox{1.2\height}{\rotatebox{90}{\parbox{\imgheight}{\centering\textbf{fp32 with cross-correlation}}}} &
    \includegraphics[width=0.18\textwidth]{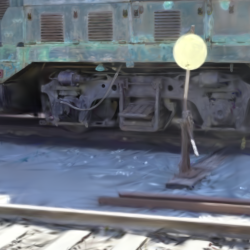} &
    \includegraphics[width=0.18\textwidth]{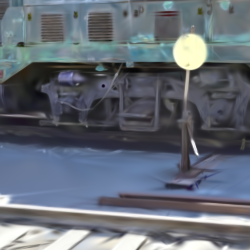} &
    \includegraphics[width=0.18\textwidth]{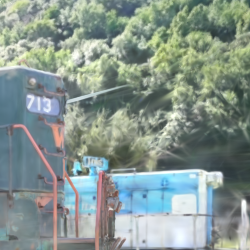} &
    \includegraphics[width=0.18\textwidth]{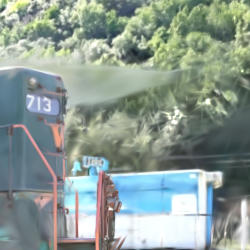} \\ [-80pt]

    % \raisebox{1.2\height}{\rotatebox{90}{\parbox{\imgheight}{\centering\textbf{fp32 without cross-correlation}}}} &
    % \includegraphics[width=0.2\textwidth]{patches/op_0.005_fp32_2.png} &
    % \includegraphics[width=0.2\textwidth]{patches/op_0.5_fp32_2.png} &
    % \includegraphics[width=0.2\textwidth]{patches/op_0.005_fp32_1.png} &
    % \includegraphics[width=0.2\textwidth]{patches/op_0.5_fp32_1.png} \\
    % [-80pt]
    
    % \raisebox{1.2\height}{\rotatebox{90}{\parbox{\imgheight}{\centering\textbf{fp16}}}}  &
    % \includegraphics[width=0.18\textwidth]{patches/op_0.005_fp16_2.png} &
    % \includegraphics[width=0.18\textwidth]{patches/op_0.5_fp16_2.png} &
    % \includegraphics[width=0.18\textwidth]{patches/op_0.005_fp16_1.png} &
    % \includegraphics[width=0.18\textwidth]{patches/op_0.5_fp16_1.png} \\
    % [-80pt]
    
    % \textbf{Int8} &
    % \includegraphics[width=0.2\textwidth]{patches/op_0.005_int8_2.png} &
    % \includegraphics[width=0.2\textwidth]{patches/op_0.5_int8_2.png} &
    % \includegraphics[width=0.2\textwidth]{patches/op_0.005_int8_1.png} &
    % \includegraphics[width=0.2\textwidth]{patches/op_0.5_int8_1.png} \\
    % % [6pt]

    \raisebox{1.5\height}{\rotatebox{90}{\parbox{\imgheight}{\centering\textbf{int8 without cross-correlation}}}}  &
        \begin{minipage}[b]{0.18\textwidth}\centering
          \includegraphics[width=\linewidth]{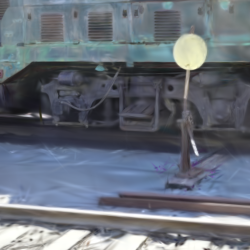}\\[2pt]
          \(\bm{\alpha=0.005}\) 
        \end{minipage} &
        \begin{minipage}[b]{0.18\textwidth}\centering
          \includegraphics[width=\linewidth]{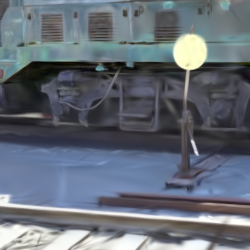}\\[2pt]
          \(\bm{\alpha=0.5}\) 
        \end{minipage} &
        \begin{minipage}[b]{0.18\textwidth}\centering
          \includegraphics[width=\linewidth]{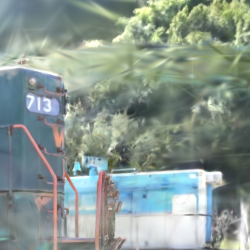}\\[2pt]
          \(\bm{\alpha=0.005}\) 
        \end{minipage} &
        \begin{minipage}[b]{0.18\textwidth}\centering
          \includegraphics[width=\linewidth]{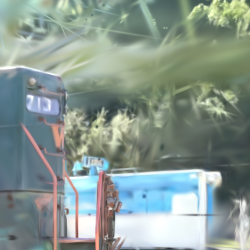}\\[2pt]
          \(\bm{\alpha=0.5}\) 
        \end{minipage} \\ [-90pt]
        
    % \textbf{Set 5} &
    % \includegraphics[width=0.23\textwidth]{example-image-a} &
    % \includegraphics[width=0.23\textwidth]{example-image-a} &
    % \includegraphics[width=0.23\textwidth]{example-image-a} &
    % \includegraphics[width=0.23\textwidth]{example-image-a} \\
    \end{tabular}

    \caption{\small Qualitative and quantitative analysis of rendering quality. The top row shows the ground truth image and the PSNR matrix. PSNR remains high even at $\alpha$=0.25, allowing for a \textbf{3.49x} reduction in 6T inverter columns. We show renders across different setups. Although the 6T array is constrained to model only diagonal covariance, this does not substantially degrade the PSNR and fidelity is preserved even with lower precision and higher $\alpha$. Lower precision reduces DAC, ADC power and area consumption.}
    \label{fig:scenes}
    \vspace{-6pt}
\end{figure*}

 Fig. \ref{main plots}(a) shows the change in variance as we sweep the threshold voltages. The range of values that the normalized variance can take is from 0.05 to 0.15. For a larger range, the multiplication operation can be done on the columnd. For instance, multiplying the value by $\beta$ changes the variance by a factor of $\beta^2$. However, in applications like gaussian splatting, the variances are typically low.

The simulations show that the variance does not change independently of the mean entirely. For the MOSFET widths ratio $W_p = 3W_n$, the deviation in mean is the least (Fig. \ref{main plots}(b)). The variations are attributed to the unsymmetrical nature of the pull-up and pull-down networks. When we change the mean, the variance remains constant, and so does the maximum current in the column. Changing the variance will affect the maximum current as the overdrive voltage changes. An ADC with a high range and precision would allow us to utilize a wide range of threshold voltages with finer bins.

\section{Energy-Performance Characterization}

We next analyze the impact of ProbSplat's hardware characteristics on scene reconstruction, illustrated in Fig. \ref{fig:scenes}. We used the train scene dataset~\cite{3dgs} to render scenes at multiple precisions and opacity thresholds ($\alpha$). We analyze of int8 and fp32 precisions for low ($\alpha=0.005$) and high ($\alpha=0.5$) opacity thresholds. Opacity threshold is the minimum opacity value to consider a point as a solid in scene reconstruction (the smaller, the better for rendering). The means and covariance matrices in gaussian splatting are programmed within the constraints of 6T FG inverter columns. Comparisons with ground truth shows promising outputs with PSNR of 21.99 dB at int8 precision with a few losses in the information without significantly disturbing the overall scene rendering quality. Information loss in the reconstructed scene increases with lower precision and higher opacity threshold as shown for the bottom images in Fig. \ref{fig:scenes}.

\begin{figure}[h]
    \centering
    \setlength{\belowcaptionskip}{-5pt}
    \begin{subfigure}[b]{0.22\textwidth}
        \centering
        \includegraphics[width=\textwidth]{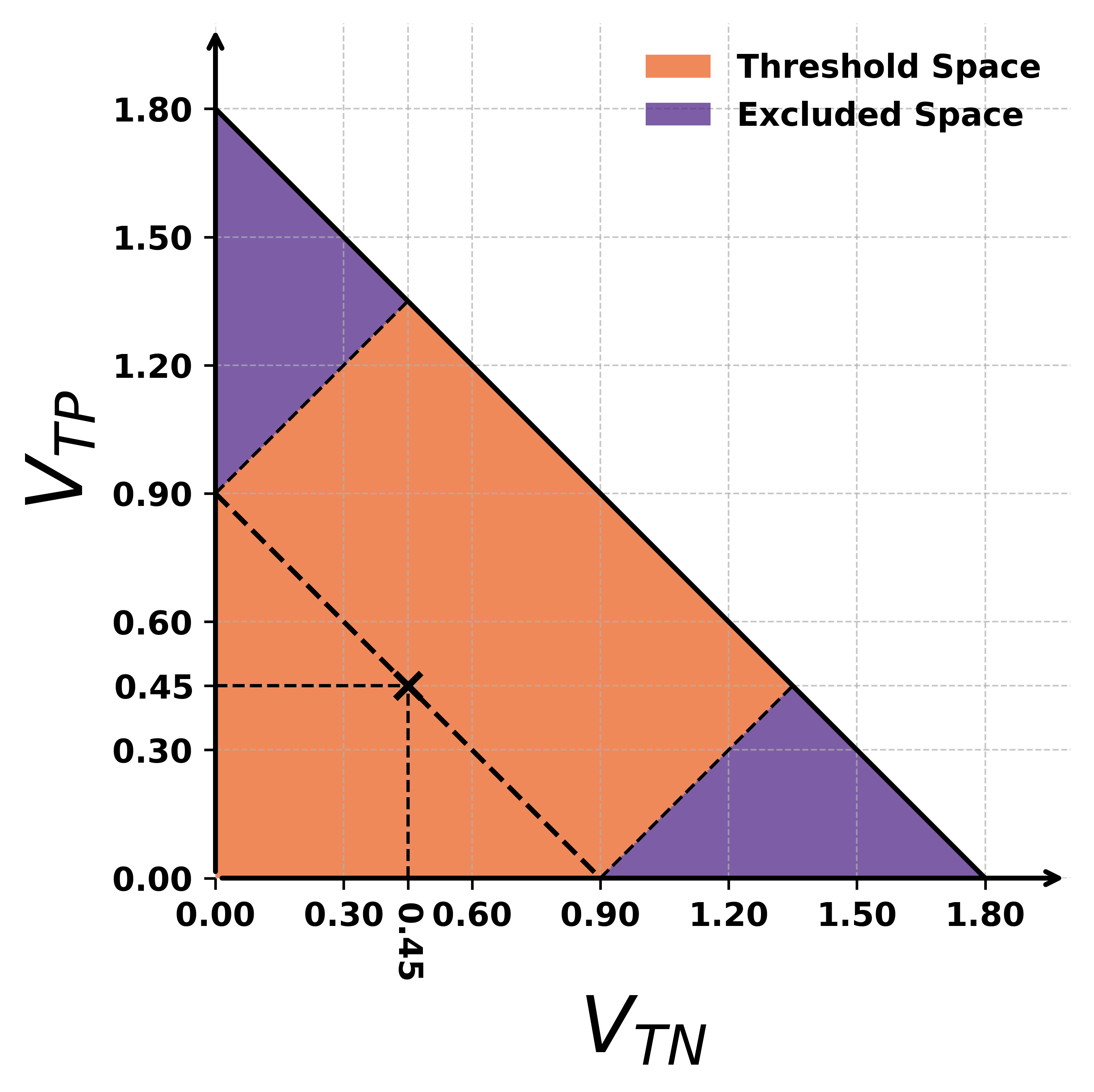} % Replace with your first plot
        \caption{}
        % \label{}
    \end{subfigure}
    \hfill % This command inserts a horizontal space between the two subfigures
    \begin{subfigure}[b]{0.22\textwidth}
        \centering
        \includegraphics[width=\textwidth]{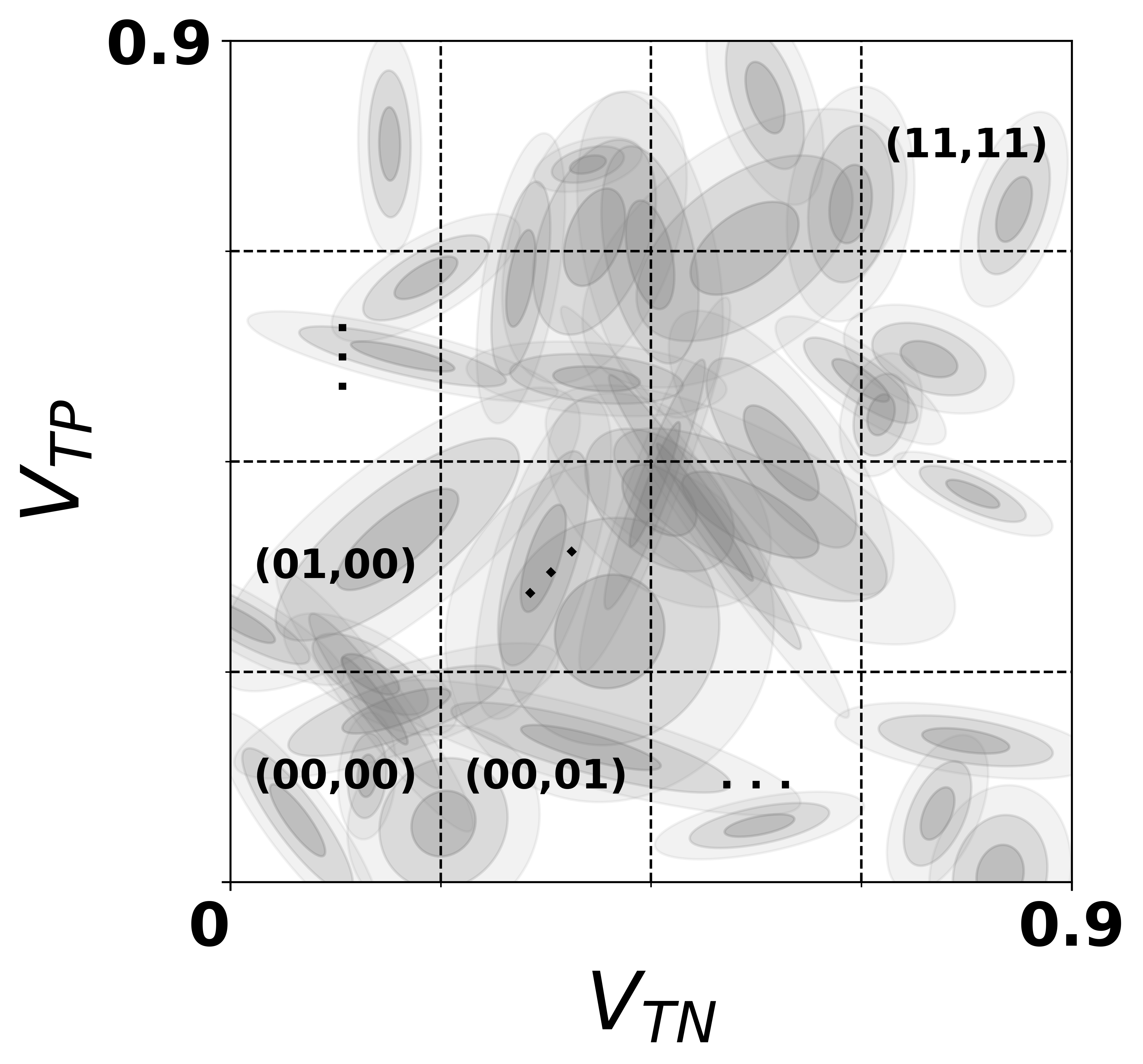} % Replace with your second plot
        \caption{}
        % \label{fig:plotB}
    \end{subfigure}
    \caption{\small{(a) Programming of a 1D Gaussian's mean and variance in a 2-D space, where the axes correspond to the threshold voltages of the respective transistors. (b) The scene's bit-encoding uses 2 bits per axis, which the control block uses to multiplex the required cells.}}
    % \vspace{-1em}
    \label{power_plots}
    \vspace{-12pt}
\end{figure}

To evaluate the energy metrics of the proposed design, 500 columns of 6T inverters were simulated, where each column was randomly assigned mean and variance values. We sample mean and variance values from a uniform distribution and find the required threshold voltages. We ran a Monte-Carlo simulation by sampling 1000 points in the range $[0, 0.9]^3$. We plot $V_{T_n}$ and $|V_{T_p}|$ on a 2-D plane to visualize the threshold voltage's effect on means and variances. The range of values $V_{T_n}$ and $|V_{T_p}|$ can take is constrained by the line \textit{x + y = 1.8} (Fig. \ref{power_plots}(a)) in order to have a region where both the MOSFETs are not turned off. Changing threshold voltages along a line with slope $-1$ will only affect the mean, and moving along a line with slope $+1$ will only affect the variance. Since the contour lines for mean and variance are orthogonal, changing one doesn't affect the other. Fig. \ref{power_plots}(a) shows the range of values the threshold voltages can take. The default values for $V_{TH}$ corresponding to the center of the scene are chosen to be 0.45 V, implying that the scene is mapped between a mean value of 0 V and 0.9 V. A higher default value would limit the variance of the gaussians at the extreme edges of the scene. On the other hand, a lower value would constrain the range of the mean.

The control block in Fig \ref{application array}(a) can activate the columns that primarily contribute to the current while deactivating the others, decreasing power consumption. Since the control block is implemented using digital logic, it incurs only dynamic power consumption and introduces negligible overhead. The simple control logic is depicted in Fig. \ref{power_plots}(b), where the scene is split into 16 cells (encoded using 2 bits per dimension). Only the gaussians in the cell where the input position belongs are activated. The peripheral ADC/DAC energy consumption is from the referenced designs~\cite{log-adc},~\cite{dac}. Under ideal energy scaling, we have assumed Energy $\propto$ (180 nm/TechNode$_{\text{REF}}$)$^2$, Energy $\propto$ (1.8 V/VDD$_{\text{REF}}$)$^2$, and Energy $\propto$ $2^{\text{Precision}_{\text{CURR}} - \text{Precision}_{\text{REF}}}$ and extracted the corresponding energy estimates. 

Table \ref{power-table} presents the projected energy of reference designs adjusted to match our design specifications. 4-bit precision and 8-bit precision suffice for several applications, like drone localization. For the 4-bit precision, the total energy per evaluation is 17.99 pJ. The Log-ADC accounts for 49.47\% of the total energy consumption, the three DACs contribute 19.01\%, and the inverter columns constitute the remaining 31.50\%. For the case of 8-bit precision, the total energy per evaluation is 202.57 pJ. The power consumed by the Log-ADC and the DACs scales accordingly. As the number of inverter columns is fixed, their energy consumption remains the same and is relatively small. This energy breakdown also indicates that as the scenes get more complex, the increase in energy would be just because of the inverter columns. The power consumed by Log-ADC/DACs remains constant. 

\captionsetup[table]{style=centeredlabel}
\begin{table}[htbp]
    \renewcommand{\arraystretch}{1.3}
    \centering
    % \captionsetup{font=small}
    \caption{\textsc{Energy Projected to 1.8V, 180}nm \textsc{CMOS, and 50MHz}}
    \label{power-table}
    \resizebox{\columnwidth}{!}{%
    \begin{tabular}{|c|c|M{1.4cm}|M{1.4cm}|}
        \hline
        \multirow{2}{*}{\textbf{Components}} & 
        \multirow{2}{*}{\textbf{Metrics (Ref.)}} & 
        \multicolumn{2}{c|}{\textbf{Energy (proposed design)}} \\ \cline{3-4}
        & & \textbf{4-bits} & \textbf{8-bits} \\ 
        \hline
        \raisebox{-.6\height}{Log-ADC} 
        & \raisebox{-.6\height}{\shortstack{2.54mW @ 1.62V, 0.18$\mu$m, \\ 8-bits, 22MS/s ~\cite{log-adc}}}
        & \raisebox{-.7\height}{8.91 pJ} 
        & \raisebox{-.7\height}{142.54 pJ} \\[1.03em]
        \hline
        \raisebox{-.7\height}{3 DACs}  
        & \raisebox{-.6\height}{\shortstack{27mW @ 1.2V, 0.13$\mu$m, \\ 10-bits, 1.6 GHz ~\cite{dac}}}
        & \raisebox{-.6\height}{3.41 pJ} 
        & \raisebox{-.6\height}{54.59 pJ} \\[1.03em]
        \hline
        \raisebox{-.6\height}{Inv columns} 
        & \raisebox{-.6\height}{-} 
        & \raisebox{-.6\height}{5.67 pJ} 
        & \raisebox{-.6\height}{5.67 pJ} \\[1.03em]
        \hline
    \end{tabular}%
    }
\end{table}

The conventional architecture in Fig. \ref{Top Fig}(b) (left) was implemented in 65-nm CMOS technology~\cite{digital-implementation}. Using power estimates of different blocks in the datapath from~\cite{power-figures} and projecting it to 180-nm CMOS, 1.8 V, the estimated energy per inference for the 8-bit GMM processor is 477 pJ for 30 3-D mixture functions. Comparatively, our proposed framework consumes 202.57 pJ at 8-bit quantization for 500 3-D mixture functions.  

\section{Conclusion}

In this work, we present a method to independently model the mean and variance of a 3-D gaussian mixture model (GMM) using 6T floating-gate inverter columns to accelerate probabilistic computing with high energy efficiency. Adopting the proposed mixed-signal in-memory (6T array) computing alleviates the use of conventional power hungry digital architecture. The proposed control scheme further enhances the energy efficiency. A detailed analysis of mean and variance programming of the 6T-inverter array is provided for Gaussian Splatting, a critical use-case in 3D scene reconstruction for augment/virtual reality (AR/VR), proving to be a promising candidate for miniaturization and energy efficiency of AR/VR devices. %However, we cannot account for the covariances between dimensions with just a single inverter column. A prospective research direction is modeling the covariance between dimensions, improving the fidelity of the scene representations.

\balance
\bibliographystyle{IEEEtran}  % or plain, alpha, unsrt, etc.
\bibliography{references}     % Do not include the .bib extension

@ARTICLE{6t-inv,
  author={Shukla, Priyesh and Muralidhar, Ankith and Iliev, Nick and Tulabandhula, Theja and Fuller, Sawyer B. and Trivedi, Amit Ranjan},
  journal={IEEE Transactions on Very Large Scale Integration (VLSI) Systems}, 
  title={Ultralow-Power Localization of Insect-Scale Drones: Interplay of Probabilistic Filtering and Compute-in-Memory}, 
  year={2022},
  volume={30},
  number={1},
  pages={68-80},
  doi={10.1109/TVLSI.2021.3100252}}

@misc{3dgs,
      title={3D Gaussian Splatting for Real-Time Radiance Field Rendering}, 
      author={Bernhard Kerbl and Georgios Kopanas and Thomas Leimkühler and George Drettakis},
      year={2023},
      eprint={2308.04079},
      archivePrefix={arXiv},
      primaryClass={cs.GR},
      url={https://arxiv.org/abs/2308.04079}, 
}

@Article{gaussian-transistors,
author={Lee, Changhyeon
and Rahimifard, Leila
and Choi, Junhwan
and Park, Jeong-ik
and Lee, Chungryeol
and Kumar, Divake
and Shukla, Priyesh
and Lee, Seung Min
and Trivedi, Amit Ranjan
and Yoo, Hocheon
and Im, Sung Gap},
title={Highly parallel and ultra-low-power probabilistic reasoning with programmable gaussian-like memory transistors},
journal={Nature Communications},
year={2024},
month={Mar},
day={18},
volume={15},
number={1},
pages={2439},
issn={2041-1723},
doi={10.1038/s41467-024-46681-2},
url={https://doi.org/10.1038/s41467-024-46681-2}
}

@article{MLprobabilistic,
  title={Probabilistic machine learning and artificial intelligence},
  author={Ghahramani, Zoubin},
  journal={Nature},
  volume={521},
  number={7553},
  pages={452--459},
  year={2015},
  publisher={Nature Publishing Group UK London}
}

@InProceedings{Charatan_2024_CVPR,
    author    = {Charatan, David and Li, Sizhe Lester and Tagliasacchi, Andrea and Sitzmann, Vincent},
    title     = {pixelSplat: 3D Gaussian Splats from Image Pairs for Scalable Generalizable 3D Reconstruction},
    booktitle = {Proceedings of the IEEE/CVF Conference on Computer Vision and Pattern Recognition (CVPR)},
    month     = {June},
    year      = {2024},
    pages     = {19457-19467}
}

@inproceedings{arvr_4dgs,
  title={4d gaussian splatting for real-time dynamic scene rendering},
  author={Wu, Guanjun and Yi, Taoran and Fang, Jiemin and Xie, Lingxi and Zhang, Xiaopeng and Wei, Wei and Liu, Wenyu and Tian, Qi and Wang, Xinggang},
  booktitle={Proceedings of the IEEE/CVF conference on computer vision and pattern recognition},
  pages={20310--20320},
  year={2024}
}

@inproceedings{fu2024colmap,
  title={Colmap-free 3d gaussian splatting},
  author={Fu, Yang and Liu, Sifei and Kulkarni, Amey and Kautz, Jan and Efros, Alexei A and Wang, Xiaolong},
  booktitle={Proceedings of the IEEE/CVF Conference on Computer Vision and Pattern Recognition},
  pages={20796--20805},
  year={2024}
}

@InProceedings{Matsuki_2024_CVPR,
    author    = {Matsuki, Hidenobu and Murai, Riku and Kelly, Paul H.J. and Davison, Andrew J.},
    title     = {Gaussian Splatting SLAM},
    booktitle = {Proceedings of the IEEE/CVF Conference on Computer Vision and Pattern Recognition (CVPR)},
    month     = {June},
    year      = {2024},
    pages     = {18039-18048}
}

@article{robotics-localization,
  title={Probabilistic terrain mapping for mobile robots with uncertain localization},
  author={Fankhauser, P{\'e}ter and Bloesch, Michael and Hutter, Marco},
  journal={IEEE Robotics and Automation Letters},
  volume={3},
  number={4},
  pages={3019--3026},
  year={2018},
  publisher={IEEE}
}

@ARTICLE{Autonomous_driving,
  author={Feng, Di and Harakeh, Ali and Waslander, Steven L. and Dietmayer, Klaus},
  journal={IEEE Transactions on Intelligent Transportation Systems}, 
  title={A Review and Comparative Study on Probabilistic Object Detection in Autonomous Driving}, 
  year={2022},
  volume={23},
  number={8},
  pages={9961-9980},
  doi={10.1109/TITS.2021.3096854}}

@article{surgical_gmm,
  title={Temporal clustering of surgical activities in robot-assisted surgery},
  author={Zia, Aneeq and Zhang, Chi and Xiong, Xiaobin and Jarc, Anthony M},
  journal={International journal of computer assisted radiology and surgery},
  volume={12},
  number={7},
  pages={1171--1178},
  year={2017},
  publisher={Springer}
}

@ARTICLE{log-adc,
  author={Lee, Jongwoo and Kang, Joshua and Park, Sunghyun and Seo, Jae-sun and Anders, Jens and Guilherme, Jorge and Flynn, Michael P.},
  journal={IEEE Journal of Solid-State Circuits}, 
  title={A 2.5 mW 80 dB DR 36 dB SNDR 22 MS/s Logarithmic Pipeline ADC}, 
  year={2009},
  volume={44},
  number={10},
  pages={2755-2765},
  doi={10.1109/JSSC.2009.2028052}}

@inproceedings{dac,
  title={A 1Tb 4b/cell 64-stacked-WL 3D NAND flash memory with 12MB/s program throughput},
  author={Lee, Seungjae and Kim, Chulbum and Kim, Minsu and Joe, Sung-min and Jang, Joonsuc and Kim, Seungbum and Lee, Kangbin and Kim, Jisu and Park, Jiyoon and Lee, Han-Jun and others},
  booktitle={2018 IEEE International Solid-State Circuits Conference-(ISSCC)},
  pages={340--342},
  year={2018},
  organization={IEEE}
}

@article{power-figures,
  title={Performance comparisons between 7-nm FinFET and conventional bulk CMOS standard cell libraries},
  author={Xie, Qing and Lin, Xue and Wang, Yanzhi and Chen, Shuang and Dousti, Mohammad Javad and Pedram, Massoud},
  journal={IEEE Transactions on Circuits and Systems II: Express Briefs},
  volume={62},
  number={8},
  pages={761--765},
  year={2015},
  publisher={IEEE}
}

@ARTICLE{digital-implementation,
  author={Price, Michael and Glass, James and Chandrakasan, Anantha P.},
  journal={IEEE Journal of Solid-State Circuits}, 
  title={A 6 mW, 5,000-Word Real-Time Speech Recognizer Using WFST Models}, 
  year={2015},
  volume={50},
  number={1},
  pages={102-112},
  doi={10.1109/JSSC.2014.2367818}}

@article{mapping_survey,
  title={Mobile robot localization: A review of probabilistic map-based techniques},
  author={Malagon-Soldara, Salvador Manuel and Toledano-Ayala, Manuel and Soto-Zarazua, Genaro and Carrillo-Serrano, Roberto Valentin and Rivas-Araiza, Edgar Alejandro},
  journal={IAES International Journal of Robotics and Automation},
  volume={4},
  number={1},
  pages={73},
  year={2015},
  publisher={IAES Institute of Advanced Engineering and Science}
}

@INPROCEEDINGS{gmm,
  author={Dasgupta, S.},
  booktitle={40th Annual Symposium on Foundations of Computer Science (Cat. No.99CB37039)}, 
  title={Learning mixtures of Gaussians}, 
  year={1999},
  volume={},
  number={},
  pages={634-644},
  doi={10.1109/SFFCS.1999.814639}}

@Article{cim,
author={Sebastian, Abu
and Le Gallo, Manuel
and Khaddam-Aljameh, Riduan
and Eleftheriou, Evangelos},
title={Memory devices and applications for in-memory computing},
journal={Nature Nanotechnology},
year={2020},
month={Jul},
day={01},
volume={15},
number={7},
pages={529-544},
issn={1748-3395},
doi={10.1038/s41565-020-0655-z},
url={https://doi.org/10.1038/s41565-020-0655-z}
}

@ARTICLE{cim-survey,

  author={Verma, Naveen and Jia, Hongyang and Valavi, Hossein and Tang, Yinqi and Ozatay, Murat and Chen, Lung-Yen and Zhang, Bonan and Deaville, Peter},

  journal={IEEE Solid-State Circuits Magazine}, 

  title={In-Memory Computing: Advances and Prospects}, 

  year={2019},

  volume={11},

  number={3},

  pages={43-55},

  doi={10.1109/MSSC.2019.2922889}}

@article{probabilistic-photonics,
  title={Probabilistic photonic computing for AI},
  author={Br{\"u}ckerhoff-Pl{\"u}ckelmann, Frank and Ovvyan, Anna P and Varri, Akhil and Borras, Hendrik and Klein, Bernhard and Meyer, Lennart and Wright, C David and Bhaskaran, Harish and Syed, Ghazi Sarwat and Sebastian, Abu and others},
  journal={Nature Computational Science},
  pages={1--11},
  year={2025},
  publisher={Nature Publishing Group US New York}
}

@article{probabilistic-approx,
  title={Probabilistic numerics and uncertainty in computations},
  author={Hennig, Philipp and Osborne, Michael A and Girolami, Mark},
  journal={Proceedings of the Royal Society A: Mathematical, Physical and Engineering Sciences},
  volume={471},
  number={2179},
  pages={20150142},
  year={2015},
  publisher={The Royal Society Publishing}
}

\end{document}